\documentclass[12pt]{article}

\usepackage{amsmath}
\usepackage{amsfonts}
\usepackage{amssymb}

\usepackage{cite}

\usepackage{graphicx}

\usepackage[hang,footnotesize]{caption2}

\usepackage[dvips]{hyperref}

\newtheorem{theorem}{Theorem}
\newtheorem{prop}[theorem]{Proposition}
\newtheorem{definition}[theorem]{Definition}
\newtheorem{remark}[theorem]{Remark}
\newtheorem{cor}[theorem]{Corollary}

\newenvironment{proof}[1][Proof]{\textbf{#1:} }{\ \rule{0.5em}{0.5em}}

\setcaptionwidth{.8\textwidth}

\newcommand{\M}{\ensuremath{\mathcal{M}}}

\begin{document}

\DeclareGraphicsExtensions{.eps}

\title{Bohmian arrival time without trajectories}

\author{Sabine Kreidl, Gebhard Gr\"ubl and Hans G Embacher\\[10pt]
Institut f\"ur Theoretische Physik der Universit\"at Innsbruck\\
Technikerstr. 25, A-6020 Innsbruck,
Austria\\[2pt]sabine.kreidl@uibk.ac.at}

\date{}

\maketitle

\begin{abstract}

The computation of detection probabilities and arrival time
distributions within Bohmian mechanics in general needs the
explicit knowledge of a relevant sample of trajectories. Here it
is shown how for one-dimensional systems and rigid inertial
detectors these quantities can be computed without calculating any
trajectories. An expression in terms of the wave function $\Psi$
and its spatial derivative $\partial_{x}\Psi$, both restricted to
the boundary of the detector's spacetime volume, is derived for
the general case, where the probability current at the detector's
boundary may vary its sign.

\end{abstract}




\section{Introduction}

A microscopic object $s$ may trigger a sudden change in the
properties of a macroscopic system $S$. Often such events take
place during an interval of time much shorter than the duration of
the interaction between $s$ and $S$. Perhaps the simplest example
is the detection time of a quantum particle which slowly passes a
detector. This \emph{phenomenologically} well accessible quantity
is usually termed arrival time. Quantum \emph{theory} has
difficulties in identifying events and a fortiori their time of
occurrence within its formalism, because, according to
Schr\"{o}dinger's equation, the unitary state evolution of the
closed system containing $s\ $and $S$ does not make any sudden
jumps. Jumps of states are introduced into the standard quantum
formalism only through state reduction, which is supposed to
happen in an open system, when a ``measurement'' is performed on
it from ``outside''. Yet in this case the instant of time of the
reduction is not stochastic but rather determined by the
observer's deliberate choice. Thus state reduction does not seem
to be the proper notion to understand the stochastic distribution
of the time of events.

Attempts to obtain the arrival time distribution through the model
of continuous observation leads to the well known quantum zeno
paradox \cite{zeno}. Also the proposals for a time operator (see
e.g. \cite{GRT}) are still subject to discussion and controversy
\cite{Muga,Muga2}.\\

One strategy to incorporate detection events into quantum theory
is by means of Bohmian mechanics (BM), which introduces the
additional notion of particle trajectories into the standard
formalism. Within this framework Leavens \cite{Leavens} and
McKinnon and Leavens \cite{McKinnon} derived an expression for the
time resolved detection probability for one-dimensional (1D)
scattering situations. Let $\Psi$ be a normalized solution of a
time dependent 1D Schr\"{o}dinger equation and $j$ its associated
probability current density. In \cite{Leavens,McKinnon} it has
been argued for a Bohmian particle with wave function $\Psi$ that
the detection probability at position $a$ between time $0$ and
time $T>0$ is given by $\int_{0}^{T}\left\vert j(t,a)\right\vert
dt$ \footnote[1]{Under certain provisos the quantum optical
detection model of reference \cite{Muga3} concurs with this
expression.}. The line of argument assumes from the outset that no
trajectory passes through $a$ more than once during the time
interval $\left[ 0,T\right]$ which is guaranteed if $j(\cdot,a)$
does not change sign. (For an ideal detector the first entry
triggers an event, e.g. by discharging the device and producing a
click. Further on the detector is insensitive to additional
entries.)

If, however, multiple crossings do occur, the replacement of
$j(t,a)$ by a cut off current has been advocated by Daumer
\emph{et al} \cite{Goldstein}, such that only the first traversal
of the trajectories should be counted. This means that the time
intervals with second, third, etc. crossings should be dropped
from the integral $\int_{0}^{T}\left\vert j(t,a)\right\vert dt$.
Therefore the computation of such detection probabilities in
general demands the explicit knowledge of the Bohmian trajectories
of the problem at hand. Yet Bohmian trajectories are the solutions
of a nonlinear system of ordinary differential equations and
therefore difficult to obtain.

The general Bohmian notion of detection probability associated
with quite arbitrary space-time regions has been formalized in
\cite{Gebhard}. As in the earlier treatment \cite{Goldstein}
Bohmian trajectories enter the formula defining the detection
probability. Here we show how for 1D systems and rigid inertial
detectors the Bohmian detection probability and the associated
arrival time distribution can be computed under quite general
circumstances without any knowledge of the Bohmian trajectories.
The relevant formula is contained in proposition \ref{mainprop}.

In section \ref{BohmM} we give a very short overview of the main
ingredients of Bohmian mechanics, omitting mathematical detail.
Section \ref{ArrTfBF} introduces the essential mathematical
structures needed for the formulation of one-dimensional arrival
time problems in the framework of non-relativistic Bohmian
mechanics. In section \ref{Main} then a reformulation of the
arrival time distribution for one-dimensional detectors occupying
spatial intervals $[a,b]$, with the mere aid of probability and
current density integrals, is given and proved. Finally section
\ref{Examples} demonstrates the practical use of the technique
given in section \ref{Main} by means of several numerical
examples. Free evolution as well as evolution under the influence
of external potentials is considered.

\section{Bohmian Mechanics}
\label{BohmM}

BM rests on the insight that with each normalized solution of the
time dependent Schr\"{o}dinger equation, a fibration of the
configuration space-time is given. At time $t$ each fibre (Bohmian
trajectory) has a unique representative in the underlying
configuration space, and therefore a dynamical evolution of
configurations along the fibres follows. The local conservation of
configuration space probability implies that its quantum
mechanical evolution coincides with the one implied by the
transport along the fibres. Therefore a causal deterministic
interpretation of quantum mechanics in terms of movement in
configuration space becomes consistent. An individual quantum
system with wave function $\Psi$ within BM is now assumed to
realize one of the system's trajectories, i.e. at each instant the
system \emph{is} in a point of the configuration space. Amending
the continuum notions of quantum theory by such point structures
opens up the possibility to identify unique properties and sudden
events within the formalism. Consequently one may ask again ``Does
a trajectory enter a certain spacetime region?'' or ``When is it,
that the system's trajectory enters a certain spacetime region?''.

Since the choice among the trajectories is beyond control, only
probabilistic predictions can be made, yet at any time a system
\emph{has} properties without the need to invoke state reduction.
Whenever the probabilistic predictions of BM can be compared with
those of standard quantum mechanics, they agree. To us the prime
achievement of BM seems to be, that it resolves the quantum
measurement problem. A concise summary of Bohmian mechanics can be
found in \cite{Goldstein2}.

\section{Arrival time from Bohmian flow}
\label{ArrTfBF} For the sake of simplicity, the set of Galilean
spacetime points is assumed to be $\M=\mathbb{R}^{2}$. As a
positively oriented, global and inertial chart we choose
$id_{\M}=:(t,x)$. The associated tangent frame is denoted by
$(\partial_t,\partial_x)$.

\subsection{Bohmian velocity vector field and Bohmian flow}

Let the mapping $\Psi: \M \rightarrow \mathbb{C}$ be
$\mathcal{C}^{2}$ and a solution to the Schr\"odinger equation
\[
\textrm{i} \hbar \partial_{t} \Psi = \left[ -\frac{\hbar^{2}}{2m}
\partial_{x}^{2} + V(x) \right] \Psi
\]
with $V$ any real scalar potential. Then with the aid of the
position and current densities
\[
\rho: = \Psi^{*}\Psi
\]
and
\[
j: =  \frac{\hbar}{m} \Im \left\{ \Psi^{*}\left(
\partial_x \Psi \right) \right\}
\]
the current vector field $\hat{j}$ is defined as
\[
\hat{j}: = \rho \partial_t + j \partial_x
\]
on \M. If $\rho \neq 0, \; \forall p \in \M$ the corresponding
Bohmian velocity vector field
\[
v: = \partial_t + \frac{j}{\rho} \partial_x
\]
is $\mathcal{C}^1$. The maximal integral curve of the vector field
$v$ through a point $p \in \M$ is the unique function $\gamma: I_p
\rightarrow \M$ with $\gamma(0)=p$ and
\[
\dot{\gamma}(s)=v(\gamma(s)), \quad \forall s \in I_p
\]
where $I_p$ is a non-extendable open real interval. Those integral
curves are regarded to represent the worldlines of the actual
Bohmian particles. We assume that $v$ is complete, which means
that $I_p = \mathbb R$, for all $p \in \M$. Then the mapping
\[
F: \mathbb{R} \times \M \rightarrow \M, \; (s,p) \mapsto
\gamma_p(s)
\]
is a global flow on \M\ with $F(r,F(s,p))=F(r+s,p)$. As $t \circ
F(s,p)=t \circ \gamma_p(s) = s+t(p)$, no worldline begins or ends
at a finite time. Moreover $F(s, \cdot)$ bijectively maps
instantaneous spaces $\Sigma_{\tau}= \{ p \in \M/ \; t(p) = \tau
\}$ onto instantaneous spaces, namely
$F(s,\Sigma_r)=\Sigma_{s+r}$.

\subsection{Conservation of probability}

\begin{figure}[h!]
\centering
\includegraphics[width=.5\textwidth]{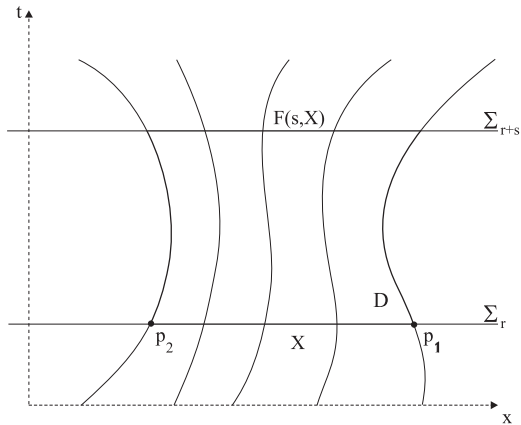}\\
 \caption{\label{Stokes1}Spacetime region $D$ enclosed by integral
curves and instantaneous sets}
\end{figure}

By inserting $\hat{j}$ as the first argument into the volume form
$E=dt \wedge dx$, we get the current-1-form
\[
J:=\hat{j} \lrcorner E = \rho \textrm{d} x -j \textrm{d} t.
\]
Due to the continuity equation $\partial_t \rho + \partial_x j=0$
the current form $J$ is closed:
\[ \textrm{d} J = (\partial_t \rho + \partial_x j) \textrm{d} t \wedge \textrm{d} x = 0.
\]
For any spacetime region $D$ with piecewise
$\mathcal{C}^{1}$-boundary $\partial D$, Stoke's theorem therefore
assures that
\begin{equation}\label{stokes}
\int_{\partial D} J = \int_D \textrm{d} J = 0.
\end{equation}
In particular for spacetime regions of the form indicated in
figure \ref{Stokes1}, equation (\ref{stokes}) for any Borel set
$X$ results in
\begin{equation}\label{CPD}
\int_{X} J = \int_{F(s,X)} J.
\end{equation}

Equation (\ref{CPD}) can be seen as follows: Let $\partial D$ be
oriented ``anticlockwise'' (direction of integration), then
\begin{eqnarray}
\int_{\partial D} J & = & \int_{X} J +
\underbrace{\int_{\gamma_{p_{1}}([r,r+s])} J}_{(a)} -
\int_{F(s,X)} J -
\underbrace{\int_{\gamma_{p_{2}}([r,r+s])}}_{(b)} J.\nonumber
\end{eqnarray}
Since
\begin{eqnarray}
\int_{\gamma_{p}([\alpha,\beta])} J & = & \int_{\alpha}^{\beta}
J(\dot{\gamma}_{p}(s)) \textrm{d} s = \int_{\alpha}^{\beta}
J(v_{\gamma_p(s)}) \textrm{d} s \nonumber \\
& = & \int_{\alpha}^{\beta} v \lrcorner(\hat{j}\lrcorner
E)_{\gamma_p(s)} \textrm{d} s = \int_{\alpha}^{\beta}
 \frac{1}{\rho} E(\hat{j},\hat{j})_{\gamma_p(s)} \textrm{d} s = 0 \nonumber
\end{eqnarray}
the contributions $(a)$ and $(b)$ to $\int_{\partial D} J$ vanish: $(a)=(b)=0$.\\

For an interval $X \subseteq \Sigma_\tau$ integrals of the form
(\ref{CPD}) expressed in our chosen coordinates read as
\begin{equation}\label{probX}
\int_X J = \int_X (\rho \textrm{d} x-j \textrm{d} t) =
\int_{\min(x(X))}^{\max(x(X))} \rho(\tau,\xi) \textrm{d} \xi
\end{equation}
which is easily recognized as the standard quantum mechanical
probability for ``finding a particle'' at time $\tau$ in the
spatial interval $[\min(x(X)),\max(x(X))]$, as soon as $\rho$ is
integrable and normalized, i.e. $\int_{-\infty}^{\infty}
\rho(\tau,\xi) \textrm{d} \xi = \int_{-\infty}^{\infty} \Psi^*
\Psi (\tau,\xi) \textrm{d} \xi = 1$, which will further on be
assumed to hold. Equation (\ref{CPD}) can then be interpreted as
the conservation of probability along the flow lines of the
Bohmian vector field. In other words, the amount of probability
contained in $X$ is the same as in $F(s,X)$, for all $s \in
\mathbb{R}$. Note that for a complete vector field $v$, we get a
fibration of \M\ into the images of the integral curves or orbits
of $v$. Equation (\ref{probX}) then induces a measure on the space
of these orbits, independently of the choice of $\tau$, which will
be argued below.

Having this in mind, the amount of probability contained in the
set of Bohmian orbits passing through quite arbitrary spacetime
regions $D$, which need not be subsets of an instantaneous space,
can be defined in a straightforward and unambiguous manner: We
simply take the probability (\ref{probX}) of the intersection of
all the flow lines, passing through $D$, with an arbitrary
hypersurface $\Sigma_{\tau}$.

Let $\pi$ be the projection $\pi: \mathbb{R} \times \M \rightarrow
\M, \;(s,p) \mapsto p$ and let $F_{\tau}$ be the restriction of
$F$ onto $\mathbb{R}\times \Sigma_{\tau}$. Then the mapping
\[
\mathcal{F}_{\tau}:=\pi \circ F_{\tau}^{-1}: \M \mapsto
\Sigma_{\tau}
\]
is the fibre projection of \M\ onto the instantaneous subspace
$\Sigma_{\tau}$ along the Bohmian trajectories. If
$\mathcal{F}_{\tau}(D)$ is a Borel set in $\Sigma_{\tau}$ then we
define the transition of the Bohmian vector field through $D$ as
\[
T(D)=\int_{\mathcal{F}_{\tau}(D)} J.
\]
With $\mathcal{X}_{\tau}:=x \circ \mathcal{F}_{\tau}$ delivering
the $x$-coordinates of the fibre projection $\mathcal{F}_{\tau}$
this yields
\[
T(D)=\int_{\mathcal{X}_{\tau}(D)} \rho(\tau,\xi) \textrm{d} \xi.
\]
Relation (\ref{CPD}) secures that $T(D)$ is independent of the
choice of the instantaneous subspace $\Sigma_{\tau}$, or
respectively of the choice of $\tau$. In what follows we will
restrict ourselves to $\mathcal{F}_{0}$ and $\mathcal{X}_{0}$
respectively.

\begin{remark}\label{Nomeasure}
$T$ does not induce a measure on the Borel subsets of \M, as
$T(D_1 \cup D_2) \neq T(D_1)+T(D_2)$ in general, for $D_1 \cap D_2
= \emptyset$. For regions $D_1 \subseteq D_2$ it is however
guaranteed, that $\mathcal{F}_{0}(D_1) \subseteq
\mathcal{F}_{0}(D_2)$, and therefore $\mathcal{X}_{0}(D_1)
\subseteq \mathcal{X}_{0}(D_2)$, as all the trajectories passing
through $D_1$ obviously also contribute to the transition through
$D_2$. I.e. for $D_1 \subseteq D_2$ the relation $T(D_1) \leq
T(D_2)$ holds.

\end{remark}

\subsection{Arrival time distribution}

\begin{figure}[h!]
\centering
\includegraphics[width=.5\textwidth]{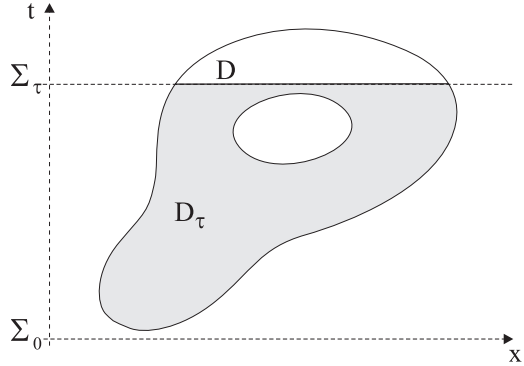}\\
\caption{Subsets $D_{\tau}$ of a spacetime region
$D$\label{Dsubsets}}
\end{figure}

Having the notion of transition through a spacetime region $D$,
the definition of detection probabilities is obtained easily. For
a spacetime region $D \subseteq \M$ we define the subsets
\begin{equation}\label{Dtau}
D_{\tau}:  = \{ p \in D/ \; t(p) \leq \tau\}
\end{equation}
as indicated in figure \ref{Dsubsets}. The transition
$T(D_{\tau})$ of flow lines through $D_{\tau}$ then is the
probability for a Bohmian particle, to have ``arrived'' in the
region $D$ before time $\tau$. If $D$ is furthermore assumed to be
the spacetime region occupied by a 100\% efficient, purely passive
detecting device, which ``clicks'' as soon as a particle, i.e.
it's Bohmian trajectory, enters it, then the transition
$T(D_{\tau})$ finds it's interpretation as detection probability
up to time $\tau$.

\begin{definition}\label{Ptau}
For a spacetime region $D$ with subsets $D_{\tau}$ (\ref{Dtau}),
whose fibre projections $\mathcal{F}_{0}(D_{\tau})$ are Borel sets
for every $\tau \in \mathbb{R}$, the detection probability
\[
P_D(\tau): = T(D_{\tau}) = \int\limits_{\mathcal{F}_{0}(D_{\tau})}
J = \int\limits_{\mathcal{X}_{0}(D_{\tau})} \rho(0,\xi) \textrm{d}
\xi
\]
is, as a function of time $\tau \in \mathbb{R}$, positive,
monotonically increasing and bounded.
\end{definition}

The positivity is guaranteed because of the positivity of the
probability density $\rho=\Psi^{*}\Psi$, the monotonicity because
of $D_{\tau_{1}} \subseteq D_{\tau_{2}}$ for $\tau_{1} \leq
\tau_{2}$ and remark \ref{Nomeasure}. $P$ is bounded because the
transition $T(D)$ is bounded by 1 for any spacetime region $D$.

Interpretatively the quantity $N:=\lim\limits_{\tau \to
\infty}P_D(\tau)$ represents the overall probability for a
detection event in $D$. Now taking only that part of Bohmian world
lines into account, which enter the spacetime region $D$ at some
time, and therefore produce a detection event, $\frac{1}{N}
P_D(\tau)$ delivers the conditional probability for a detection
event up to time $\tau$.

\begin{definition}\label{ATdist}
Let $P_D$ be the detection probability function of definition
\ref{Ptau}, then by the function
\[
P_D^c:=\frac{P_D}{\lim\limits_{\tau \to \infty} P_D(\tau)}:
\mathbb{R} \rightarrow [0,1]
\]
the conditional arrival time distribution for a spacetime region
$D$ shall be denoted.
\end{definition}

The lower index of $P_{D}$ and $P_{D}^c$ respectively will further
on be omitted, if it is evident from the context, which spacetime
region $D$ is meant.

\section{Formulation without trajectories \label{Main}}

From now on we consider detectors occupying spacetime regions of
the form
\begin{equation}\label{det}
D=\{ p \in \M / x(p) \in [a,\, b],\; t(p) \geq 0 \}
\end{equation}
which corresponds to a detector being at rest with respect to our
inertial chart, occupying the interval $[a,\, b]$, and being
sensitive from time $t=0$ onward. For $b \to a$ we get the point
detector at $x=a$.

Again we have the subsets
\[
D_{\tau}=\{ p \in D/ \;t(p) \leq \tau \}.
\]

\begin{prop}\label{mainprop}

For the detection probability function $P$ for a spacetime region
of type (\ref{det}) the formula
\begin{equation}\label{mainequ}
P(\tau)=P(0)+\max \{ f_a(s)/ \; 0 \leq s \leq \tau\} + \max \{
-f_b(s)/ \; 0 \leq s \leq \tau\}
\end{equation}
holds for times $\tau \geq 0$, with
\[
f_a(s)=\int_{0}^{s} j(t,a) \textrm{d} t \qquad \textrm{and} \qquad
f_b(s)=\int_{0}^{s} j(t,b) \textrm{d} t
\]
being antiderivatives of the current densities at $x=a$ and $x=b$,
respectively.

\end{prop}

\begin{proof}
\noindent From remark \ref{Nomeasure} we know that
\[
P(\tau)=T(D_{\tau})=\int\limits_{\min(\mathcal{X}_{0}(D_{\tau}))}^{\max(\mathcal{X}_{0}(D_{\tau}))}
\rho(0,\xi) \textrm{d} \xi \; \geq \;
P(0)=T(D_0)=\int\limits_{a}^{b} \rho(0,\xi) \textrm{d} \xi
\]
and also that
\[
\min(\mathcal{X}_{0}(D_{\tau})) \leq a \qquad \textrm{and} \qquad
\max(\mathcal{X}_{0}(D_{\tau})) \geq b
\]
as $\mathcal{X}_{0}(D_0)=[a,\,b] \subseteq
\mathcal{X}_{0}(D_{\tau})$, which allows the separation
\begin{eqnarray}\label{Ptausplit}
P(\tau) & = & \int\limits_{\min(\mathcal{X}_{0}(D_{\tau}))}^{a}
\rho(0,\xi) \textrm{d} \xi + \int\limits_{a}^{b} \rho(0,\xi)
\textrm{d} \xi + \int\limits_{b}^{\max(\mathcal{X}_{0}(D_{\tau}))}
\rho(0,\xi) \textrm{d} \xi
\nonumber \\
& = & \int\limits_{\min(\mathcal{X}_{0}(D_{\tau}))}^{a}
\rho(0,\xi) \textrm{d} \xi + P(0) +
\int\limits_{b}^{\max(\mathcal{X}_{0}(D_{\tau}))} \rho(0,\xi)
\textrm{d} \xi.
\end{eqnarray}
The non-crossing property of Bohmian trajectories secures that
\[
\mathcal{X}_{0}(\{(t,a)\}) \leq \mathcal{X}_{0}(\{(t,x)\}), \quad
\forall \; x \geq a
\]
and
\[
\mathcal{X}_{0}(\{(t,x)\}) \leq \mathcal{X}_{0}(\{(t,b)\}), \quad
\forall \; x \leq b.
\]
This allows the reformulation of (\ref{Ptausplit}) into

\begin{eqnarray}
P(\tau) & = & P(0) \nonumber\\
& + &   \int_{\min(\mathcal{X}_{0}(D_{\tau}^{a}))}^{a} \rho(0,\xi)
\textrm{d} \xi
\label{t1}  \\
& + & \int_{b}^{\max(\mathcal{X}_{0}(D_{\tau}^{b}))} \rho(0,\xi)
\textrm{d} \xi \label{t2}
\end{eqnarray}

with
\[
D_{\tau}^{a}=\{ p \in D_{\tau}/ \;x(p)=a \}
\]
and
\[
D_{\tau}^{b}=\{ p \in D_{\tau}/ \;x(p)=b \}
\]
being the restrictions to the right, respectively left, edges of
$D_{\tau}$. Because of the positivity of the position density
$\rho$, by applying Stoke's theorem to regions as illustrated in
figure \ref{Stokes2}, term (\ref{t1}) can be expressed as
\begin{eqnarray}
\int_{\min(\mathcal{X}_{0}(D_{\tau}^{a}))}^{a} \rho(0,\xi)
\textrm{d} \xi & = & \max\limits_{0 \leq t \leq \tau } \left\{
\int_{\mathcal{X}_{0}(\{(t,a)\})}^{a}
\rho(0,\xi) \textrm{d} \xi \right\}\nonumber\\
= \max\limits_{0 \leq t \leq \tau} \left\{ \int_{0}^{t} j(s,a)
\textrm{d} s \right\} & = &  \max \left\{ f_{a}(t) / \; 0 \leq t
\leq \tau \right\}. \label{Stja}
\end{eqnarray}

\begin{figure}[h!]
\centering
\includegraphics[width=.4\textwidth]{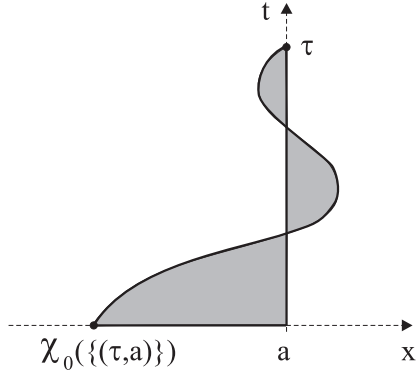}\\
\caption{Spacetime region\label{Stokes2}}
\end{figure}

Analogously term (\ref{t2}) becomes
\begin{eqnarray}
\int_{b}^{\max(\mathcal{X}_{0}(D_{\tau}^{b}))} \rho(0,\xi)
\textrm{d} \xi & = & \max\limits_{0 \leq t \leq \tau} \left\{
\int_{b}^{\mathcal{X}_{0}(\{(t,b)\})}
\rho(0,\xi) \textrm{d} \xi \right\} \nonumber\\
= \max\limits_{0 \leq t \leq \tau} \left\{ \int_{t}^{0} j(s,b)
\textrm{d} s \right\} & = & \max \left\{ -f_{b}(t)/ \; 0 \leq t
\leq \tau \right\}. \label{Stjb}
\end{eqnarray}
Equation (\ref{Ptausplit}) together with (\ref{Stja}) and
(\ref{Stjb}) finally reads
\[
P(\tau)=P(0)+\max \{ f_a(s)/ \; 0 \leq s \leq \tau\} + \max \{
-f_b(s)/ \; 0 \leq s \leq \tau\}
\]
which completes the proof.
\end{proof}\\

Taking the limit $b \to a$ we immediately get the detection
probability for point detectors:

\begin{cor}

For spacetime regions of the form
\[
D= \{ p \in \M / \; x(p)=a,\, t(p) \geq 0\}
\]
the detection probability function $P$ can be expressed as
\[
P(\tau)=\max \{ f_{a}(s)/ \; 0 \leq s \leq \tau \}+\max \{
-f_{a}(s)/ \; 0 \leq s \leq \tau \}
\]
with
\[
f_a(s)=\int_{0}^{s} j(t,a) \textrm{d} t
\]
being an antiderivative of the current density at $x=a$.
\end{cor}

The conditional arrival time distribution associated with the
spacetime region (\ref{det})
\[
P^{c}:=\frac{P}{\lim\limits_{s \to \infty}P(s)}: \mathbb{R}
\rightarrow [0,1]
\]
obeys $P^{c}(\tau)=0$ for $\tau<0$. $P^{c}(0)>0$ holds for $a<b$,
since $\rho > 0$. For $\tau >0$ $P^c$ is continuous and $P^c$ is
non-decreasing. Thus $P^c$ is the distribution function of a
Lebesgue-Stieltjes measure $\mu$ \cite{Loeve}. $\mu$ can be
separated into a point measure $\mu_s$, which ascribes to every
Borel set $A \subseteq \mathbb{R}$ the value $P^{c}(0)$ if $0 \in
A$, the value 0 otherwise, and an absolutely continuous part
$\mu_{ac}$, with respect to the Lebesgue measure. That is, there
exists a Lebesgue-measurable density function $\delta$ on
$\mathbb{R}$, which vanishes for $\tau \leq 0$ and ascribes to
every interval $[a,b]$ the value
\[
\mu_{ac}([a,b])=P^{c}(b)-P^{c}(a)=\int_a^b \delta(s) \textrm{d} s
\]
if $0 < a < b$. $\mu$ finally reads as $\mu=\mu_s+\mu_{ac}$

The density $\delta$ is needed for the calculation of the
expectation values and variances of the arrival time, represented
by the stochastic variable
\[
T^{A}: \mathbb{R} \to \mathbb{R}, \tau \mapsto \tau.
\]
If the relevant integrals exist, they can be obtained in the usual
manner:
\[
\langle T^{A} \rangle_{\mu} = \langle T^{A}
\rangle_{\mu_s}+\langle T^{A} \rangle_{\mu_{ac}} = 0 \cdot P^c(0)+
\int_{0}^{\infty} \tau \cdot \delta(\tau)  \textrm{d} \tau  =
\int_{0}^{\infty} \tau \cdot \delta(\tau) \textrm{d} \tau
\]
and
\[
\mathcal{V}_{\mu}(T^{A}) = \langle (T^{A})^2 \rangle_{\mu} -
\langle T^{A} \rangle_{\mu}^2.
\]

\begin{remark}

The probability density $\delta$ on $]0,\infty[$ takes the form
\begin{eqnarray}\label{PrDensity}
\delta(\tau)& = &\left( \lim\limits_{t \rightarrow \infty} P(t)
\right) ^{-1} \cdot \left[ j(\tau,a) \cdot
\Theta\left(f_a(\tau)-\max\limits_{0 \leq s \leq \tau}\{ f_a(s) \}
\right) \right.
\nonumber\\
& - &  \left. j(\tau,b) \cdot
\Theta\left(-f_b(\tau)-\max\limits_{0 \leq s \leq \tau}\{ -f_b(s)
\} \right) \right]
\end{eqnarray}
where $\Theta$ denotes the step function $\Theta(s)=\left\{
\begin{array}{cc} 0 & s<0
\\ 1 & s \geq 0
\end{array} \right.$.

The formulation for a point detector at $x=a$ is simply achieved
by replacing $b$ with $a$ everywhere in (\ref{PrDensity}).
\end{remark}

This shows, that it is not enough to know the current density at a
given instant $\tau$, rather the current density $j(\cdot,a)$ has
to be known at all instants within the interval $[0, \tau]$, in
order to compute the probability density $\delta$ of the arrival
time distribution at $a$. Note however, that the Bohmian
trajectories do not need to be known. The function $\delta \cdot
\lim\limits_{t \rightarrow \infty} P(t)$ is the cut off current
introduced in \cite{Goldstein}.

\section{Examples}\label{Examples}

The following numerical examples shall give an overview of the
applicability of our treatment.

\subsection{Free evolution}

\begin{figure}[h!]
\centering
\includegraphics[width=0.8\textwidth
]{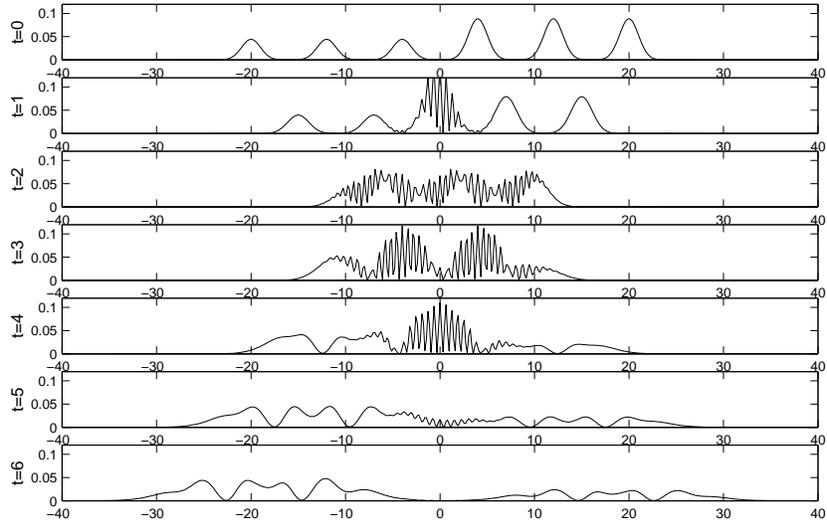}\\
\caption{Free evolution of a one-particle wavefunction consisting
of six Gaussians, moving to the right and left, respectively.
\label{Psifrei}}
\end{figure}

As a first example we choose a solution $\Psi$ to the free and
parameter-reduced Schr\"odinger equation
\[
\textrm{i} \partial_t \Psi = -\frac{1}{2} \partial_x^2 \Psi
\]
The parameter reduction is simply achieved by taking the
$t$-coordinates in units of $\frac{m}{\hbar q^2}$ and the
$x$-coordinates in units of $1/q$, where $q$ is a characteristic
wave number of the wavefunction (e.g. corresponding to a peak in
the momentum space). We choose $\Psi$ to be of the form
\begin{eqnarray}
\Psi(t,x) & = & \sqrt{\frac{1}{9}} \big( \Phi(t,x;k_0,x_0)+
\Phi(t,x;k_0,3 x_0)+ \Phi(t,x;k_0,5 x_0)\big) \nonumber\\ & + &
\sqrt{\frac{2}{9}} \big( \Phi(t,x;-k_0,-x_0)+ \Phi(t,x;-k_0,-3
x_0)+ \Phi(t,x;-k_0,-5 x_0)\big) \nonumber
\end{eqnarray}
with
\[
\Phi(t,x;k_0,x_0)=\left( \frac{d^2}{2 \pi} \right)^{1/4}
\frac{\exp(-k_0^2 d^2)}{\sqrt{d^2+\textrm{i} t/2}} \exp \left(
\frac{(2 d^2 k_0+\textrm{i} (x-x_0))^2}{4 d^2+ 2 \textrm{i} t }
\right)
\]
and consequently initial data $\Psi_0=\Psi(0,\cdot)$. This
describes six differently weighted Gaussian wave packets moving
with the same velocity in opposite directions. The evolution of
this wave function for the chosen parameters $k_0=5$, $d=1$ and
$x_0=-4$ is illustrated in figure \ref{Psifrei}.

\begin{figure}[h!]
\centering
\includegraphics[width=0.8\textwidth
]{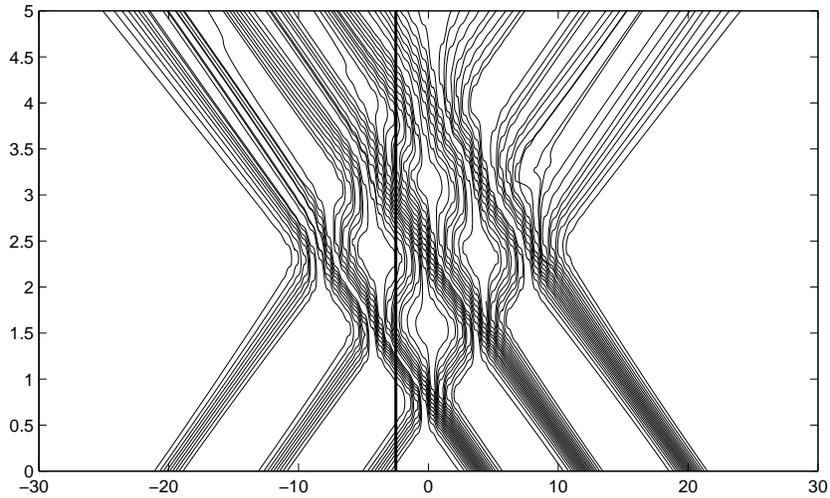}\\
\caption{Bohmian trajectories \label{Trajfrei}}
\end{figure}

The corresponding Bohmian trajectories, beeing solutions to the
velocity vector field $v$, show the peculiar non-intersection
property (figure \ref{Trajfrei}). Even ``free'' particles change
their direction of motion along their paths. The starting points
of the worldlines are sampled according to $\vert \Psi_0 \vert^2$.

\begin{figure}[h!]
\centering
\includegraphics[width=0.8\textwidth
]{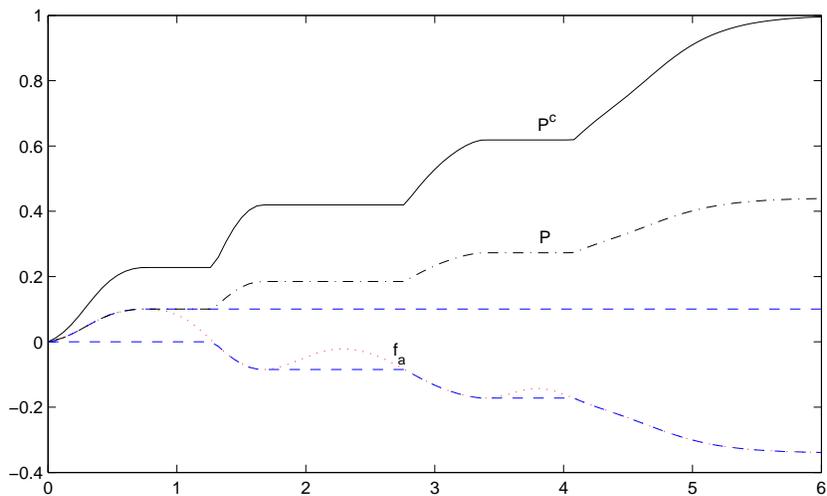}\\
\caption{$f_a$ (dotted line), $\max \{ f_a(s) / \; 0 \leq s \leq t
\}$, $-\max \{ -f_a(s)/ \;  0 \leq s \leq t \}$ (upper and lower
dashed lines), the detection probability function $P$
(dashed-dotted line) and the conditional arrival time distribution
$P^c$ (solid line).\label{AZSfrei}}
\end{figure}

We place a point detector $D$ at $x=-2.5$. The corresponding
arrival time distribution is gained with the method of proposition
\ref{mainprop}. It shows areas of non-increasing arrival time
probability, which correspond to times, during which already
detected particles enter D for a second, third, etc. time, and
therefore do not contribute to $P(t)$ anymore. Figure
\ref{AZSfrei} illustrates this phenomenon and the technical
procedure of proposition \ref{mainprop}.

\subsection{The potential barrier}

Terms as arrival-, delay- or dwell-times, etc in the literature
are often connected to scattering states of one-dimensional
potential barriers. For reviews of the subject see
\cite{Hauge,Recami1,Recami2}. We apply equation (\ref{mainequ}) to a situation like that.\\

Now the evolution of our wavefunction $\Psi$ is given by the
Schr\"odinger equation
\[
\textrm{i} \partial_t \Psi = \left[ -\frac{1}{2}  \partial_x^2 +
\frac{1}{2} \big( \Theta(x-a)-\Theta(x-b) \big) \right] \Psi
\]
with $a<b$ and $\Theta$ denoting the step function. The parameter
reduction in this case can be achieved by taking the
$t$-coordinates in units of $\frac{\hbar}{2 V_0}$ and the
$x$-coordinates in units of $\frac{\hbar}{\sqrt{2 m V_0}}$, where
$V_0$ is assumed to be the height of the potential barrier.

\begin{figure}[h!]
\centering
\includegraphics[width=0.8\textwidth
]{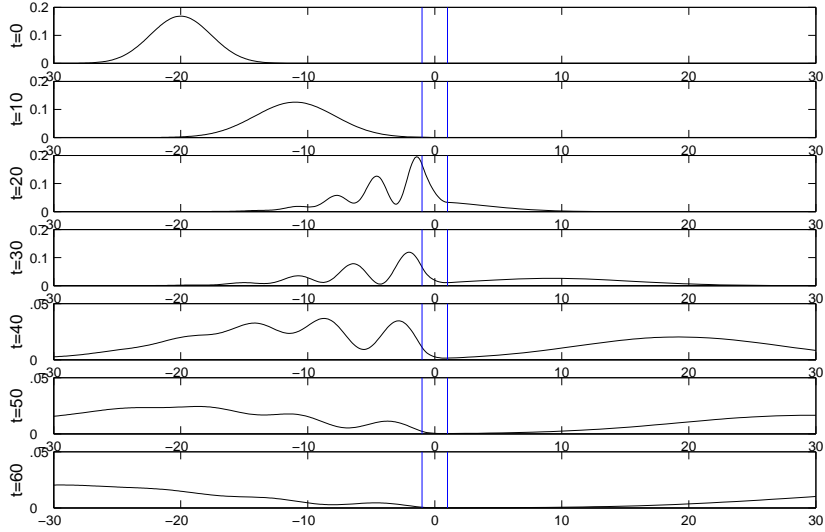}\\
\caption{Evolution of a Gaussian wave packet impinging on a
potential barrier.\label{PsiqPotB}}
\end{figure}

\begin{figure}[h!]
\centering
\includegraphics[width=0.8\textwidth
]{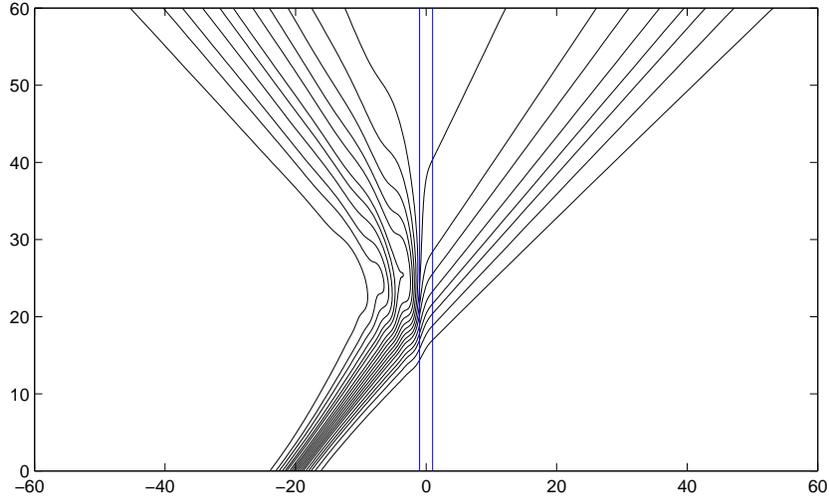}\\
\caption{Bohmian trajectories \label{TrajPotB}}
\end{figure}

The initial gaussian wave packet $\Psi_0$ is placed sufficiently
far to the left of the potential barrier (for the purpose that at
time $t=0$ the interference effects due to the barrier are
negligible). $\Psi$ is moving towards the barrier. The evolution
of the wavefunction is illustrated in figure \ref{PsiqPotB}.
Figure \ref{TrajPotB} shows the corresponding Bohmian trajectories
sampled according to $\vert \Psi_0 \vert^2$.

\begin{figure}[h!]
\centering
\includegraphics[width=0.8\textwidth
]{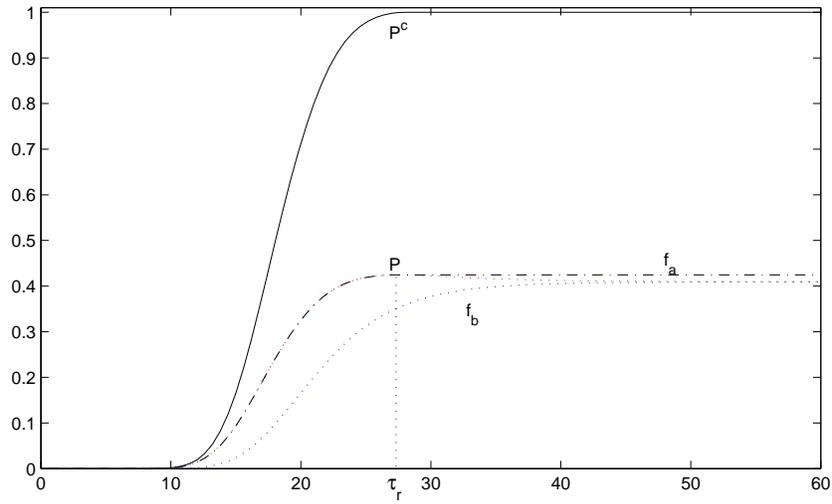}\\
\caption{$f_a$ and $f_b$ (upper and lower dotted lines), the
detection probability function $P$ (dashed-dotted line) and the
arrival time distribution $P^c$ (solid line) for the Gaussian wave
packet impinging on a potential barrier. \label{AZSPotB}}
\end{figure}

We consider an extended detector occupying the spacetime region of
the barrier from time $t=0$ onwards. As the wave packet is placed
sufficiently far to the left of the barrier, the probability of a
detection event at $t=0$ is negligible. Figure \ref{AZSPotB}
shows, that $f_b$ is positive and monotonically increasing, which
corresponds to a positive current density at the right edge of the
potential barrier. Therefore $f_b$ does not contribute to an
increase of the detection probability. As trajectories enter the
detector from the left, $f_a$ at the left edge of the barrier
increases up to a time $\tau_r$ when the first trajectory is
reflected before entering the detection region. From $\tau_r$
onwards then part of the trajectories inside the barrier return to
be finally reflected and therefore produce a negative current
density at $x=a$, which leads to a small decrease of $f_a$. The
resulting detection probability function $P$ and the conditional
arrival time distribution $P^c$ are indicated in figure
\ref{AZSPotB} by the dashed-dotted and solid lines, respectively.

\subsection{The potential step}

\begin{figure}[h!]
\centering
\includegraphics[width=0.77\textwidth
]{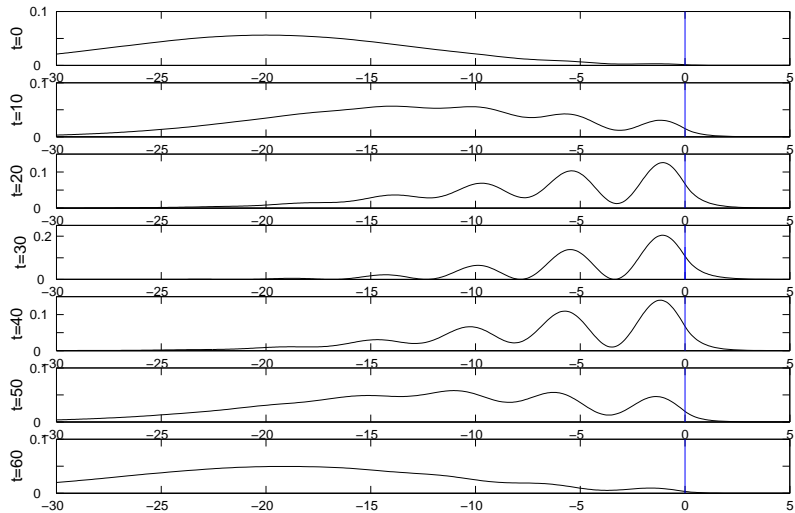}\\
\caption{Evolution of a Gaussian wave packet being totally
reflected at a potential step at $x=0$.\label{PsiqPotS}}
\end{figure}

As a third numerical example we take the case of total reflection
at a potential step at $x=0$. The wavefunction $\Psi$ is now the
solution of the Schr\"odinger equation
\[
\textrm{i} \partial_t \Psi =  \left[ -\frac{1}{2} \partial_x^2 +
\frac{1}{2} \Theta(x) \right] \Psi
\]
with $\Theta$ again denoting the step function. The parameter
reduction can be achieved analogously to the case of the potential
barrier. Again the initial data $\Psi_0$ of $\Psi$ is taken to be
a Gaussian wave packet placed to the left of the potential step at
$t=0$ (figure \ref{PsiqPotS}).

\begin{figure}[h!]
\centering
\includegraphics[width=0.8\textwidth
]{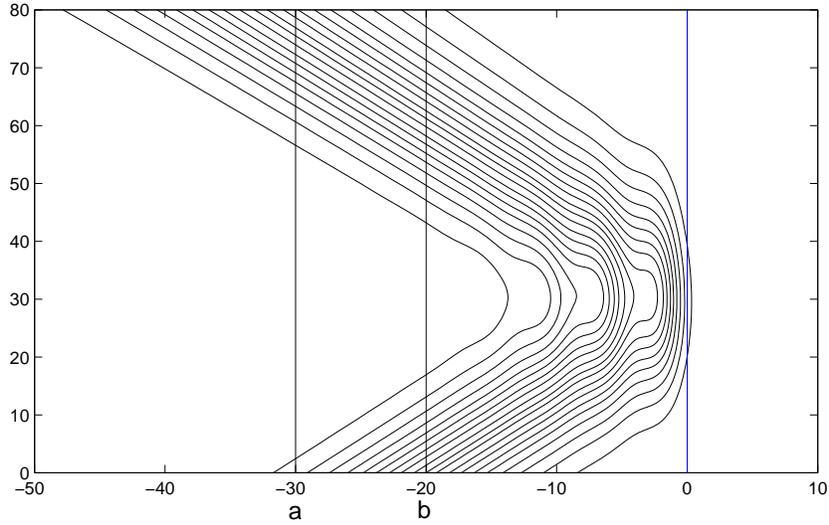}\\
\caption{Bohmian trajectories \label{TrajPotS}}
\end{figure}

This time we consider the situation, that at the time the detector
is activated, some of the trajectories are already located inside
the detection region $D$, located between $x=a$ and $x=b$ in front
of the potential step, as illustrated in figure \ref{TrajPotS}.

\begin{figure}[h!]
\centering
\includegraphics[width=0.8\textwidth
]{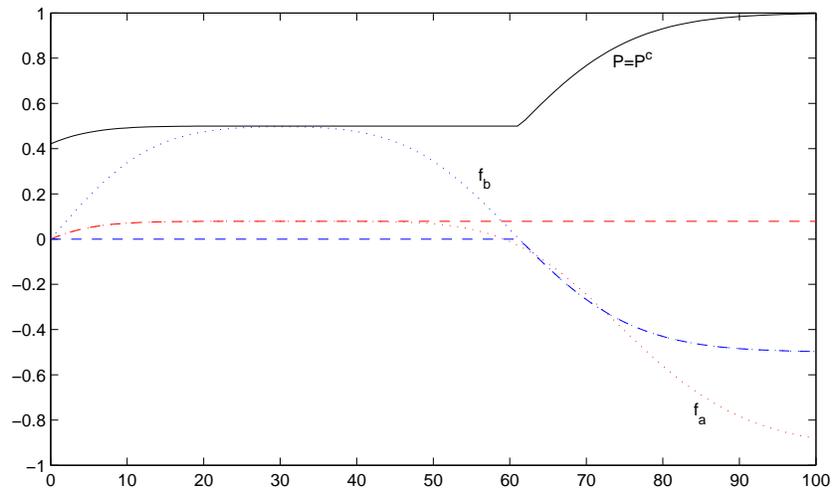}\\
\caption{$f_a$ and $f_b$ (dotted lines), $\max \{ f_a(s) / \; 0
\leq s \leq t \}$ and $-\max \{ -f_b(s)/ \;  0 \leq s \leq t \}$
(upper and lower dashed lines), the detection probability function
$P$ and the conditional arrival time distribution $P^c$ (solid
line).\label{AZSPotS}}
\end{figure}

The detection probability at time $t=0$ thus takes a value
noticeably different from 0. As we have the case of total
reflection, all the trajectories, which initially started to the
right of $b$, eventually turn back and pass the detector at a
later instance. Therefore almost all the trajectories pass $D$ at
some time and the limit $\lim\limits_{s \to \infty} P(s)$ is
approximately 1. The detection probability function and the
conditional arrival time distribution thus coincide and are
indicated by the solid line in figure \ref{AZSPotS}.

\end{document}